\documentclass[runningheads,dvipsnames,table]{llncs}
\usepackage[normalem]{ulem}
\usepackage{balance}
\usepackage{bold-extra}
\usepackage{syntax}
\usepackage{syntax}
\usepackage{color}
\usepackage{hyperref}
\usepackage{listings}
\usepackage{tikz}
\usepackage{xspace}
\usepackage{float}
\usepackage{placeins}
\usepackage[font=small,labelfont=bf]{caption}
\usepackage{wrapfig}
\usepackage{inconsolata}
\usepackage{url}
\usepackage{graphicx}
\usepackage{mdframed}
\usepackage[english]{babel}
\usepackage{blindtext}
\usepackage{stmaryrd}
\usepackage{eulervm}
\usepackage{booktabs}
\usepackage{mathtools}
\usepackage{amssymb}
\usepackage[capitalise]{cleveref}
\usepackage{algorithm}
\usepackage{algpseudocode}
\usepackage{xargs} 
\usepackage[colorinlistoftodos,prependcaption]{todonotes}
\usepackage[utf8]{inputenc}
\usepackage{tikz}
\usetikzlibrary{snakes,arrows,shapes,positioning,calc,fit}
\usepackage{amsmath}
\usepackage{pifont}
\usepackage{multirow, makecell}
\usepackage{cite}
\usepackage{subfig}
\usepackage{capt-of}

\definecolor{vgreen}{RGB}{104,180,104}
\definecolor{vblue}{RGB}{49,49,255}
\definecolor{vorange}{RGB}{255,143,102}
\RequirePackage{commands}

\crefname{section}{\S}{\S}

\lstdefinestyle{verilog-style}
{
    language=Verilog,
    basicstyle=\footnotesize\ttfamily,
    keywordstyle=\color{vblue},
    identifierstyle=\color{black},
    commentstyle=\color{vgreen},
    numbers=left,
    numberstyle=\tiny\color{black},
    extendedchars=true,
    numbersep=10pt,
    tabsize=4,
    moredelim=*[s][\colorIndex]{[}{]},
    literate=*{:}{:}1,
    xleftmargin=5.0ex,
    captionpos=b,
    escapechar=\$,
    escapeinside={(*}{*)}
}

\usepackage{xcolor}

\begin{document}

\title{\papertitle}
\author{
{\rm Rami Gökhan Kıcı \inst{1}}
\and
{\rm Klaus v. Gleissenthall \inst{2}}
\and
{\rm Deian Stefan \inst{1}}
\and
{\rm Ranjit Jhala \inst{1}}
} 
\institute{University of California San Diego \and Vrije Universiteit Amsterdam}
\date{}
\maketitle

\thispagestyle{empty}


\mypara{Abstract} 
We present \Tool, a solver-aided method for formally verifying 
that Verilog hardware executes in constant-time.
\Tool scales to realistic hardware designs by drastically reducing the effort
needed to localize the root cause of verification failures via a new notion of
{constant-time counterexamples}, which \Tool uses to automatically {synthesize} a
minimal set of secrecy assumptions. \Tool further exploits modularity in
\Verilog code via a notion of {module summaries}, thereby avoiding duplicate
work across multiple module instantiations.
We show how \Tool's assumption synthesis and summaries
enable the verification of a variety of circuits including
\AES, a highly modular AES-256 implementation where 
modularity cuts verification from six hours to 
under three seconds, and \ScarV, a timing channel hardened RISC-V
micro-controller whose size exceeds previously verified designs by an order of
magnitude.
%


\section{Introduction}

Timing side-channel attacks are no longer theoretical curiosities.
Over the last two decades, they have been
used to break implementations of cryptographic primitives ranging from
public-key encryption algorithms~\cite{kocher1996timing, brumley2005remote, cachebleed}, 
to block ciphers~\cite{bernstein2005cache, osvik2006cache},
digital signature schemes~\cite{tpmfail}, zero-knowledge
proofs~\cite{cachequote}, and pseudorandom generators~\cite{blackswan}.
This, in turn, has allowed attackers to break systems that rely on these
primitives for security---for example, to steal TLS keys used to encrypt web
traffic~\cite{brumley2005remote, cachebleed, blackswan}, to snoop and forge
virtual private network traffic~\cite{tpmfail}, and to extract information from
trusted execution environments~\cite{cachequote, stacco, blackswan,
brasser2017software}.
%
%

The gold standard for preventing timing side-channel attacks 
is to follow a discipline of \emph{constant-time} or \emph{data-oblivious} 
programming~\cite{barthe:ct, bearssl, yu2019data, binsec, ctverif, fact}.
At its core, this discipline ensures that \textbf{(1)} secret data
is not used as an operand to variable-time instructions (\eg floating point
operations like division~\cite{ctfp, raccoon, andrysco2015, kohlbrenner2017})
and \textbf{(2)} the program's control flow and
memory access patterns do not depend on secrets.

%
But for the constant-time discipline to be effective,   
it is crucial that the constant-time property be preserved by
the underlying hardware. 
For example, an instruction that is deemed 
constant-time needs to indeed produce its outputs after the 
same number of clock cycles, irrespective of operands or internal state. 
Similarly, given that control-flow and memory access patterns are \emph{public}, 
\ie free of secrets, a CPU's timing must indeed be secret-independent.   
 
Unfortunately, simply \emph{assuming} that hardware preserves constant-time doesn't work.
Incorrect assumptions about the timing-variability of floating-point
instructions, for example, allowed attackers to break the differentially
private Fuzz database~\cite{fuzz:usec11}.
Attempts to address these attacks (\eg~\cite{escort}) were also foiled: they
relied on yet other incorrect microarchitectural assumptions (\eg about the
timing-variability of SIMD instructions)~\cite{kohlbrenner2017}.
Yet more recently, hardware crypto co-processors (\eg Intel and
STMicroelectronics's trusted platform modules) turned out to exhibit similar
secret-dependent timing variability~\cite{tpmfail}.

%
A promising path towards eliminating such attacks is to 
\emph{formally verify} that our hardware preserves the constant-time 
property of the software it is executing.
%
%
Such verification efforts, however, require tool support.
Unfortunately, unlike software verification of
constant-time, which has had a long history~\cite{sok-cac}, constant-time
hardware verification is still in its infancy~\cite{yu2019data, iodine,
ct-foundations}.
As a result, existing verification approaches like \Iodine~\cite{iodine} fail 
to scale to realistic hardware.
This has two fundamental reasons.
First, existing tools do not help when verification fails,
as is inevitable, since hardware circuits only preserve constant-time execution 
under very specific \emph{secrecy assumptions} that describe which 
port and wire values are public or secret.
Currently, upon failure, the user must undertake the tedious and time consuming 
task of determining, whether the circuit is indeed leaky (\ie
variable-time), or whether it is missing additional secrecy assumptions, 
which the user then has to explicate. In our experience, this process takes up 
the overwhelming majority of time spent on the verification effort.
Second, current methods fail to exploit the \emph{modularity} 
that is already explicit at the register transfer level.
Hence, they duplicate verification effort across replicated modules which leads
to a blow up in verification time. 

In this paper, we present \Tool, a solver-aided method for formally verifying 
that \Verilog hardware executes in constant-time.
We develop \Tool via four contributions.
%
  
\mypara{1. Counterexamples}
To help users understand verification failures, we introduce the notion of
constant-time \emph{counterexamples} (\Cref{sec:min-cex}).
A counterexample highlights the \emph{earliest} point in the circuit where
timing variability is introduced; this simplifies the task of understanding
whether a circuit is variable-time by narrowing the user's attention to the
root cause of the verification failure (and a thereby small fraction of the circuit).
To compute counterexamples, \Tool leverages information extracted 
from the failed proof attempt. In particular, the solver communicates \textbf{(1)} 
for which variables (\ie registers and wires) it was able to prove the 
constant-time property and \textbf{(2)} in which \emph{order} the remaining 
variables became non-constant time. This allows \Tool to break cyclic 
data-dependencies which cause a chicken-and-egg problem that is hard to resolve 
when assigning blame manually.

\mypara{2. Assumption Synthesis}
To help the user resolve the verification failure,
\Tool uses the counterexample to synthesize a suggested fix. 
For example, \Tool may find a constant-time counterexample for a processor 
pipeline where the two different runs may execute two different ISA instructions 
(say, addition and division) which take different numbers of clock cycles. 
Yet, the execution of each instruction (for any inputs) may be constant-time.
\Tool uses the counterexample to synthesize a minimal candidate set of secrecy assumptions
(\eg that any two executions have the same, publicly visible sequence 
of instructions) which address the root cause of the verification 
failure (\Cref{sec:cex-assum}).
The user then decides either to accept the candidate assumption, or, if they do
not match their intuition for the intended usage of the circuit, reject them, in
which case \Tool computes an alternative.
Internally, \Tool computes candidate assumptions, via a reduction to integer linear programming~\cite{LIP}.

\mypara{3. Modular Verification}
Finally, in order to scale both verification and counterexample generation to
larger and more complex hardware, we introduce a notion of module summaries
(\Cref{section:vcgen}).
Module summaries succinctly capture the timing properties of a module's input and
output ports at a given usage site.
By abstracting inessential details about the exact computations performed by
the module and focusing solely on its timing behavior, 
\Tool produces fewer and more compact constraints. 

\mypara{4. Evaluation}
We implement \Tool and evaluate the impact of counterexamples, assumption
synthesis and modularity on the verification of different kinds of hardware
modules (\Cref{sec:evaluation}).
We find that \Tool's counterexample synthesis dramatically reduces the number of
potential error locations a user has to inspect in order to understand the
root-cause of the violation (to 6\% of its original size) and that \Tool is
highly effective at suggesting useful assumptions---on average 81.67\% are
accepted by the user.
Similarly, our evaluation shows that module summaries are key to reducing
verification times for certain hardware designs (\eg for AES-256 crypto core
summaries reduced the verification time from six hours to a three seconds).
Overall, we find that \Tool's solver-aided verification process drastically
reduces verification effort (\eg manually verifying the largest
benchmark of \cite{iodine} took us several minutes instead of
days) and is key to scaling verification to realistic
hardware: the SCARV side-channel hardened RISC-V core~\cite{scarv-cpu} we
verify is an order of magnitude larger than the RISC-V cores verified by
previous state-of-the-art tools.

\section{Overview}
We start by reviewing how to specify and verify the absence of timing channels
in Verilog hardware designs (\Cref{sec:overview-verif}), show how existing
techniques fail to scale on real-world hardware designs, as these designs are
often only constant-time under additional secrecy assumptions which are tedious
to derive by hand (\Cref{sec:overview-cex}), sketch how \Tool helps finding
secrecy assumptions automatically (\Cref{sec:overview-secrecy}), and finally
discuss how \Tool exploits modularity (\Cref{sec:overview-modularity}).

\subsection{Verifying Constant-Time Execution of Hardware \label{sec:overview-verif}}
\mypara{Lookup Circuit}
\Cref{fig:lookup} shows the code for a \Verilog module, which implements a
lookup table by case-splitting over the 8-bit input value. This module executes
in constant-time: even if input \var{in} contains a secret value, producing
output \var{out} takes the same amount of time (one clock cycle), irrespective of
the value of \var{in}, and therefore, an attacker cannot make any inference
about the value of~\var{in} by observing the timing of the computation.

\mypara{Specifying Constant-Time Execution}
\Cref{fig:runs-lookup} makes this intuition more precise, using a
recent definition of constant-time execution for hardware~\cite{iodine}.
\begin{wrapfigure}[13]{r}{.63\textwidth}
  \centering
  \resizebox{.4\textwidth}{!}{
  %
\begin{lstlisting}[style={verilog-style}]
module S (clk, in, out);
   input clk; [7:0] in;
   output reg [7:0] out;
   always @ (posedge clk)
     case (in)
       8'h00: out <= 8'h63;
       ...
       8'hff: out <= 8'h2c;
     endcase
endmodule
\end{lstlisting}
  %
  %
  }
  \caption{A simple, constant-time lookup-table in \Verilog, taken from~\cite{tiny-aes}. \label{fig:lookup}}
\end{wrapfigure}
Instead of tracking timing indirectly through information
flow~\cite{secverilog,li2011caisson,Kasnter11} the definition uses a direct
notion of timing.\footnote{This definition has the advantage of being more
precise. In fact, indirect techniques would \emph{not} be able to prove
constant-time execution of this simple example, since there is an information
flow from the potentially sensitive input to the output.} 
The Figure shows two runs of module~$S$: one for input \verb|8'h00| and
one for input~\verb|8'hff|. We want to track how long it takes for the two
inputs, issued at cycle~$1$ to pass through the circuit and produce their
respective outputs. 
\begin{figure}[t]
    \centering
    \centering
    \footnotesize
      \begin{tabular}{ | c ||
        >{\centering}p{0.5cm} | >{\centering}p{0.5cm} |
        >{\centering}p{0.5cm} | >{\centering}p{0.5cm} |
        >{\centering}p{0.5cm} | >{\centering}p{0.5cm} |
        >{\centering}p{0.5cm} | >{\centering}p{0.5cm} | }
        \hline
        \multirow{2}{*}{time} &
        \multicolumn{2}{c|}{\var{in}} &
        \multicolumn{2}{c|}{\var{out}} &
        \multicolumn{2}{c|}{\colorOf{\var{in}}} &
        \multicolumn{2}{c|}{\colorOf{\var{out}}} \cr \cline{2-9}
  
        & L & R & L & R & L & R & L & R \cr \hline
  
        $1$ &
        \verb|h00| & \verb|hff| &
        \UnknownValue & \UnknownValue &
        \activeCol & \activeCol &
        \deadCol & \deadCol \cr \hline
  
        $2$&
        \verb|h00| & \verb|hff| &
        \verb|h63| & \verb|h2c| &
        \deadCol & \deadCol &
        \activeCol & \activeCol \cr \hline
      \end{tabular}
  \caption{Two runs of \Cref{fig:lookup} showing values and liveness-bits input
    (\var{in}) and output (\var{out}). \UnknownValue
  \xspace represents an undefined value.\label{fig:runs-lookup}}
  \end{figure}
For this, we put a ``tracer'' on the inputs by assigning a \emph{liveness-bit} to
each register. For some register~\var{x}, we set its liveness-bit
\colorOf{\var{x}} to \activeCol, if \var{x} has been influenced by the input at
initial cycle~$1$ (we say \var{x} is \emph{1-live}) and \deadCol, otherwise.
\Cref{fig:runs-lookup} shows how liveness-bits are propagated through the circuit.
Initially, in both executions, the input is $1$-live and the output is not. In
cycle~$2$, both outputs $1$-live due to the
case-split on the value of~\var{in}. Assuming that an attacker can observe the
liveness-bits of all outputs, here, register~\var{out}, the attacker cannot
distinguish the two executions, and we can conclude that the pair of executions
is indeed constant-time.

\mypara{Verifying Constant-Time Execution}
In order to show constant-time execution, not only for the two runs in
\Cref{fig:runs-lookup}, but for the whole circuit, \Tool proves that for
\emph{any} pair of runs, that is, for any pair of inputs, and any initial cycle,
the constant-time property holds.
%
\Tool's solver achieves this by constructing a \emph{product
circuit}~\cite{iodine} whose runs correspond to \emph{pairs} of runs---called
\emph{left} and \emph{right}---of the original circuit. In this product, each
original variable \var{x} has two copies $\leftOf{\var{x}}$ and
$\rightOf{\var{x}}$ that hold the values of \var{x} in the left and right runs,
respectively.
Following \cite{iodine}, \Tool's solver uses the product circuit to synthesize
invariant properties of the circuit. For example, let's define that a
variable~\var{x} is constant time (and write $\ctVar{x}$), if for any pair of
executions, its liveness-bit in the left execution $\leftOf{\colorOf{\var{x}}}$ is
always the same as its liveness-bit in the right execution
$\leftOf{\colorOf{\var{x}}}$, \ie $\leftOf{\colorOf{x}}=\rightOf{\colorOf{x}}$
always holds, for all initial cycles~$t$. \Tool synthesizes the
following invariant on the module which proves constant-time execution, under
the condition that module inputs are constant-time: $\ctVar{in} \Rightarrow
\ctVar{out}$.

\begin{wrapfigure}[34]{l}{.5\textwidth}
  \vspace{-4ex}
  \footnotesize	
  \centering
  \resizebox{.6\textwidth}{!}{
\begin{lstlisting}[style={verilog-style}]
// source(IF_pc); sink(WB_reg);  (* \label{code:ppln-annot}       *)
module mips_pipeline(clk,rst);
input clk, rst;

assign IF_pc4 = IF_pc + 32'd4;  (* \label{code:ppln-stall-logic}       *)
assign IF_pcj = ID_Jmp  ? ID_jaddr : IF_pc4;
assign IF_pcn = M_PCSrc ? M_btgt   : IF_pcj;

assign ID_rs = ID_instr[25:21];
assign ID_rt = ID_instr[20:16];
rom32 IMEM(IF_pc, IF_instr);

always @((***))             (* \label{code:ppln-stall-bit}       *)
  Stall = EX_MemRead &&
          ( EX_rt == ID_rs ||
            EX_rt == ID_rt );
always @(posedge clk)
  if (rst)                   (* \label{code:ppln-rst}       *)
    ID_instr   <= 0; ...
  else
    if (Stall == 1) begin      (* \label{code:ppln-stall-chk} *)
      ID_instr <= ID_instr;    (* \label{code:ppln-stall}     *)
      IF_pc    <= IF_pc;
      EX_rt    <= EX_rt;
      ...
      WB_reg <= WB_reg;
    end else begin
      ID_instr <= IF_instr;    (* \label{code:ppln-fwd}       *)
      IF_pc    <= IF_pcn;      (* \label{code:advc-pc}       *)
      EX_rt    <= ID_rt;
      ...
      WB_reg <= WB_wd;
    end
  end
\end{lstlisting}
  }
  \caption{MIPS Pipeline Fragment.}
  \label{fig:pipeline}
\end{wrapfigure}
\subsection{Real World Hardware is Not Constant-Time \label{sec:overview-cex}}
Unfortunately, unlike the simple lookup-table from \Cref{fig:lookup}, real world
circuits are typically \emph{not} constant-time, in an absolute sense.
Instead, when carefully designed, they are constant-time under specific
\emph{secrecy assumptions} detailing which circuit inputs are supposed to be
\emph{public} (visible to the attacker) or \emph{secret} (unknown to the
attacker).
Thus, verification requires the user to painstakingly discover
secrecy assumptions through manual code inspection, which can be
prohibitively difficult in real world circuits.

\mypara{A Pipelined MIPS Processor}
We illustrate the importance of these assumptions using the program in
\Cref{fig:pipeline} which shows a a code-fragment taken from one of our
benchmarks---a pipelined MIPS processor~\cite{MIPS}.
If the reset bit \var{rst} is set, the processor sets a number of registers to
zero (\cref{code:ppln-rst}). Otherwise, the processor checks whether the
pipeline is stalled (\cref{code:ppln-stall-chk}) and either forwards the current
instruction from the instruction-fetch stage to the instruction-decode stage
(\cref{code:ppln-fwd}) and advances the program counter (\cref{code:advc-pc}),
or stalls by reassigning the current values (\cref{code:ppln-stall}).

\mypara{The Pipeline is not Constant-Time}
When using the processor in a security critical context, we want to make sure
that it avoids leaking secrets through timing, \ie that it is constant-time.
Unfortunately, our example pipeline is \emph{not} constant-time without any
further restrictions on its usage. For example, the execution time for a given
instruction depends on whether or not the pipeline is stalled before the
instruction is retired.
This is illustrated in \Cref{fig:run}. We model an attacker that can measure how
long an instruction takes to move through the pipeline, \ie from \emph{source}
\var{IF_pc} to \emph{sink} \var{WB_reg} (this is specified through the
annotation in \cref{code:ppln-annot} of \Cref{fig:pipeline}).
Such an attacker can distinguish the two runs in \Cref{fig:run}, as the
liveness-bits of \var{WB_reg} differ in cycle~$3$.
This timing difference lets the attacker make inferences about
the control-flow of the program which is executed on the processor, 
and therefore any attempt to verify constant-time execution results in a failure.

%

%
\begin{figure}[t]

  \centering
  \small
    \begin{tabular}{ | c ||
      >{\centering}p{0.7cm} |
      >{\centering}p{0.7cm} |
      >{\centering}p{0.7cm} |
      >{\centering}p{0.7cm} |
      >{\centering}p{0.7cm} |
      >{\centering}p{0.7cm} |
      >{\centering}p{0.7cm} |
      >{\centering}p{0.7cm} | 
      >{\centering}p{0.7cm} | 
      >{\centering}p{0.7cm} | 
      >{\centering}p{0.7cm} | 
      >{\centering}p{0.7cm} | 
      }

      \hline

      \multirow{2}{*}{time}
      & \multicolumn{2}{c|}{\var{stall}}
      & \multicolumn{2}{c|}{\var{ID_jmp}}
      & \multicolumn{2}{c|}{\colorOf{\var{IF_inst}}}
      & \multicolumn{2}{c|}{\colorOf{\var{ID_inst}}}
      & \multicolumn{2}{c|}{\colorOf{\var{EX_rt}}}
      & \multicolumn{2}{c|}{\colorOf{\var{WB_reg}}} \cr \cline{2-13}

      & L & R & L & R & L & R & L & R & L & R & L & R \cr \hline

       $1$ &
        0 & 1 &
        1 & 0 &
       \activeCol & \activeCol &
       \deadCol & \deadCol &
       \deadCol & \deadCol &
       \deadCol & \deadCol \cr \hline 
       $2$ &
        0 & 0 &
        0 & 1 &
        \deadCol & \activeCol &
        \activeCol & \deadCol &
       \deadCol & \deadCol &
       \deadCol & \deadCol \cr \hline
       $3$ &
        0 & 0 &
        0 & 0 &
        \activeCol & \deadCol &
        \activeCol & \activeCol &
        \activeCol & \deadCol &
        \cellhighlight \activeCol & \cellhighlight \deadCol \cr \hline


    \end{tabular}
  %
  \caption{Two runs of \Cref{fig:pipeline}, where the right run stalls in
    cycle~$1$. The liveness bits of sink $\var{wb_reg}$ differ in cycle~$3$ and
    therefore the circuit is not constant-time.}
  \label{fig:run}
\end{figure}

\subsection{Automatically Finding Secrecy Assumptions \label{sec:overview-secrecy}}
We may, however, still be able to use this processor safely, if we can find a
suitable set of \emph{secrecy assumptions}.
For example, we could assume that register \var{Stall} is \emph{public} (\ie
free of secrets, which can be formally expressed by the assumption that
$\leftOf{\var{Stall}}=\rightOf{\var{Stall}}$ always holds).
In this case, the timing difference in \Cref{fig:run} would only leak
information that the attacker is already aware of.
However, \var{Stall} is defined deep inside the pipeline and therefore, it is
hard to translate this assumption into a restriction on the kind of
\emph{software} \ie code we are allowed to execute on the processor.
Instead, we want to phrase our assumptions in terms of \emph{externally visible}
computation inputs. For example, restricting program counter~\var{IF_pc} to be
public directly translates into the obligation that the executed program's
control-flow be independent of secrets.

To achieve this using existing technology, the engineer first has to
\emph{manually} identify that the timing variability is first introduced in
variable \var{ID_instr} (\cref{code:ppln-fwd}) due to a control dependency on
\var{Stall}.
They then need to inspect how \var{Stall} is set (\cref{code:ppln-stall-bit})
and painstakingly trace the definitions which may involve complex combinatorial
logic (excerpt starting in line \cref{code:ppln-stall-logic}) and circular
data-flows in order to identify a promising candidate register, such that
marking the register as \emph{public} will render the circuit constant-time.
Counterexamples like \Cref{fig:run} are often of little help as they are hard to
interpret, and fail to focus attention on the relevant parts of the circuit. 

\mypara{Solver Aided Verification: \Tool's Interactive Loop}
\Tool drastically simplifies this time consuming process through a solver-aided
workflow that helps finding secrecy assumptions automatically.

\mypara{Step 1}
First, we start with an \emph{empty} set of secrecy assumptions and run \Tool on
the pipeline.
The verification fails, as the pipeline is not constant-time, however, \Tool
displays the following prompt to guide the user towards a solution.
\begin{verbatim}
> Mark 'reset' as PUBLIC? [Y/n]
\end{verbatim}
The user either says \verb+Y+ indicating that \var{reset}
should indeed be considered public, or else responds \verb+N+
which tells \Tool to \emph{exclude} the variable from future
consideration (\ie not suggest it in future).
Suppose that we follow \Tool's advice, and click \verb+Y+:
this marks \var{reset} public and re-starts \Tool for another
verification attempt.

\mypara{Step 2}
Next, \Tool suggests marking \var{M_PCSrc}
as public. \var{M_PCSrc} acts as a flag that indicates whether
the current instruction in the memory stage contains an indirect
jump.
But since \var{M_PCSrc}'s value depends on register values (\ie \var{M_PCSrc}
is set depending on whether the output of the execute phase is zero) assuming
that \var{M_PCSrc} is public would lead to assumptions about the \emph{data-memory}
which we wish to avoid. We therefore tell \Tool to exclude it in future
verification attempts and restart verification.

\mypara{Step 3}
Restarting verification causes \Tool to suggest candidate variables
\var{IF_pc} and \var{IF_pcn}, the program counter of the fetch stage
and its value in the next cycle, respectively.
We mark \var{IF_pc} as public as this directly encodes the assumption that the
program's control flow does not depend on secrets. In addition, \Tool infers a
second set of usage assumptions which details the parts of the pipeline that
have to be flushed on context switches. These assumptions would otherwise have
to be supplied by the user as well. Finally, \Tool proves that the resulting
program executes in constant-time and therefore concludes the verification
process.

\mypara{Counterexamples}
While synthesizing candidate assumptions, \Tool internally computes a
constant-time \emph{counterexample} which consists of the set of variables that
have lost the constant-time property earliest. While the user can simply follow
\Tool's suggestions without further investigating the root cause of the
violation, we find that---if the user chooses to do so---the counterexample
often helps to further understand \emph{why} the circuit has become non
constant-time. For example in our simple pipeline, \Tool returns a
counterexample consisting only of variable~$\var{ID\_instr}$ for all three
interactions. 
Indeed, inspecting the parts of the circuit where
$\var{ID\_instr}$ is assigned focuses our attention on the relevant parts of the
circuit, that is, the conditional assignment of $\var{ID\_instr}$  under
$\var{rst}$ (\cref{code:ppln-rst}) and under $\var{Stall}$
(\cref{code:ppln-stall-chk}).
We discuss how \Tool computes counterexamples using an artifact extracted from
the failed proof attempt in \Cref{sec:min-cex}, and how \Tool synthesizes secrecy
assumptions via a reduction to integer linear programming in
\Cref{sec:cex-assum}.

\subsection{Real World Circuits Aren't Not Small \label{sec:overview-modularity}}

While \Tool's solver-aided verification loop significantly reduces the time the
user has to spend on verification efforts, large real world circuits often
also present a challenge for to solver. This is because computing invariants
requires a \emph{whole program} analysis. Hence, the efficiency of our solver
crucially depends of the \emph{size} of the circuit we are analyzing, and
therefore verification might become prohibitively slow on large designs.

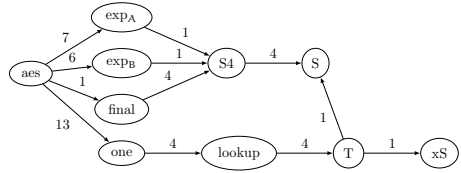
\begin{wrapfigure}[12]{r}{.5\textwidth}
%
    \centering
    \resizebox{.5\textwidth}{!}{

\begin{tikzpicture}[>=latex,line join=bevel,node distance=0.4cm and 1.4cm,auto]
  \node (u9) [draw, ellipse,inner sep=4pt]                    {$\text{aes}$};   
  \node (u5) [draw, ellipse, below right=of u9,inner sep=4pt]  {$\text{final}$}; 
  \node (u4) [draw, ellipse, below=of u5,inner sep=4pt]        {$\text{one}$};   
  \node (u7) [draw, ellipse, above=of u5,inner sep=4pt]       {$\text{exp}_B$};  
  \node (u8) [draw, ellipse, above=of u7,inner sep=4pt]       {$\text{exp}_A$};  
  \node (u3) [draw, ellipse, right=of u4,inner sep=4pt] {$\text{lookup}$};
  \node (u0) [draw, ellipse, right=of u3,inner sep=4pt]       {$\text{T}$};     
  \node (u2) [draw, ellipse, right=of u0,inner sep=4pt]       {$\text{xS}$};    
  \node (u6) [draw, ellipse, right=of u7,inner sep=4pt]       {$\text{S4}$};    
  \node (u1) [draw, ellipse, right=of u6,inner sep=4pt]       {$\text{S}$};     

  \draw [->] (u0) -- (u1) node [midway] {1};
  \draw [->] (u0) -- (u2) node [midway] {1};
  \draw [->] (u3) -- (u0) node [midway] {4};
  \draw [->] (u4) -- (u3) node [midway] {4};
  \draw [->] (u5) -- (u6) node [midway] {4};
  \draw [->] (u6) -- (u1) node [midway] {4};
  \draw [->] (u7) -- (u6) node [midway] {1};
  \draw [->] (u8) -- (u6) node [midway] {1};
  \draw [->] (u9) -- (u4) node [midway, below left=0.05cm] {13};
  \draw [->] (u9) -- (u5) node [midway] {1};
  \draw [->] (u9) -- (u7) node [midway, right=0.1cm, above=0.05cm] {6};
  \draw [->] (u9) -- (u8) node [midway] {7};
\end{tikzpicture}

}
    \caption{
    Module dependency graph of the AES-256 benchmark.
    }
    \label{fig:aes-module-dep}
%
\end{wrapfigure}

Consider, for example, the AES-256 benchmark from \cite{tiny-aes}.
\cref{fig:aes-module-dep} depicts the dependency graph of its modules, where
each node $m$ represents a Verilog module, and we draw an edge between modules
$m$ and $n$, if $m$ instantiates $n$. Each edge is annotated with the number of instantiations.
Even though there are only ten modules,
the total number of module instantiations is 789. This, in turn, causes a blowup
in the size of code \Tool has to verify.
Even though the sum of \#LOC of the modules is only 856, inlining module
instances causes this number to skyrocket to 135194 rendering verification all
but intractable. (In fact, \Tool does manage to verify the naive, inlined circuit, however,
verification takes over 6 hours to complete).

Fortunately, we can avoid this blowup by exploiting the modularity that is
already apparent at the \Verilog level.
We illustrate this process using module \var{S} from \Cref{fig:lookup}.

\mypara{Module Summaries}
Since the value of \var{out} only depends on \var{in}, we can characterize its
timing behavior as follows: the module output
\var{out} is constant-time, if module input \var{in} is constant-time.
\begin{figure}[t]
  \centering
  \resizebox{.7\linewidth}{!}{
\begin{lstlisting}[style={verilog-style}]
module S4 (clk, in, out);
   input clk;
   input [31:0] in;
   output [31:0] out;
	 wire [7:0]    out_0, out_1, out_2, out_3;
   S S_0 (clk, in[31:24], out_3),
     S_1 (clk, in[23:16], out_2),
     S_2 (clk, in[15:8],  out_1),
     S_3 (clk, in[7:0],   out_0);
	 assign out = {out_3, out_2, out_1, out_0};
endmodule
\end{lstlisting}
  }
  \caption{Module from the AES benchmark. \label{fig:module-instantiation}}
  \vspace{-4ex}
\end{figure}

We can formalize this in the following module summary, which
we \Tool computes automatically: $\ctVar{in} \Rightarrow \ctVar{out}$.
Instead of inlining the module, we can now use its summary thereby eschewing the
code explosion and enabling efficient verification.
The code in \Cref{fig:module-instantiation} shows an instantiaton of module
\verb|S| in module \verb|S4|.
Instead of inlining \verb|S| at its four instantiaton sites \verb|S_0| to
\verb|S_3|, \Tool uses the single module summary to compute a correctness proof,
which only takes 3 seconds.
\section{Modular Constant-Time Verification}
\label{section:vcgen}
We now formalize the concepts introduced in the overview. We first review a
formal definition of constant-time execution for hardware circuits
(\Cref{sec:vcgen-def}) and its translation to Horn-clause verification
conditions (\Cref{sec:vcgen-cons}). We then show how to modularize this naive
encoding via summaries (\Cref{sec:vcgen-sum}).

\subsection{Defining Constant-Time Execution \label{sec:vcgen-def}}

\mypara{Configurations}
Configurations represent the state of a \Verilog computation.
A configuration $\Sigma \eqdef (P, \sigma,\theta, c, t, \textsc{Src})$ is made
up of a \Verilog program $P$ (say, the processor in \Cref{fig:pipeline}), a
store $\sigma$, a liveness map~$\theta$, current clock cycle~$c \in \mathbb{N}$,
initial clock cycle~$t \in \mathbb{N}$ and, finally, a set of
sources~$\textsc{Src} \subseteq \textsc{Vars}$. Store $\sigma \in \textsc{Vars}
\to \mathbb{Z}$ maps variables~$\textsc{Vars}$ (registers and wires) to their
current values; map~$\theta \in \textsc{Vars} \to \theSet{\activeCol, \deadCol}$ maps
variables to liveness-bits; cycle $t$ marks the
starting-cycle of the computation we want to track and finally $\textsc{Src}$ identifies
the inputs of the computation we are interested in.

\mypara{Transition relation}
Transition relation~$\rightsquigarrow \in \left ( \Sigma \times \Sigma \right )$
encodes a standard Verilog semantics which defines how a configuration is
updated from one clock-cycle to the next. We omit its definition, as it is not
needed for our purposes, but formal accounts can be found in
\cite{gordon95,iodine,secverilog}. In addition to updating the store and current
cycle, the transition relation updates the liveness map~$\theta$ by tracking
which variables are currently influenced by the computation started in~$t$. At
initial cycle $t$, our transition relation starts a new computation by setting
the liveness-bits of all variables in $Src$ to $\activeCol$, and those of all
others variables to $\deadCol$.

\mypara{Runs}
We call a sequence of configurations $\pi \eqdef \Sigma_0 \Sigma_1 \dots
\Sigma_{n-1}$ a \emph{run}, if each consecutive pair of configurations is related by
the transition relation, \ie if $\Sigma_i \rightsquigarrow \Sigma_{i+1}$, for $i
\in \theSet{0, \dots, n-2}$.
We call $\Sigma_0 \eqdef (P, \sigma_0, \theta_0, 0, t,  \textsc{Src})$ initial state,
and require that $\theta_0$ maps all variables to~$\deadCol$.
Finally, for a run $\pi \eqdef (P, \sigma_0,\theta_0, c_0, t, \textsc{Src})
\dots (P, \sigma_{n-1},\theta_{n-1}, c_{n-1}, t, \textsc{Src})$, we say that
$\pi$ is a run of $P$ of length~$n$ with respect to~$t$ and $\textsc{Src}$ and let
$\mathit{store}(\pi,i)=\sigma_i$ and $\mathit{live}(\pi,i) \eqdef \theta_i$, for
$i \in \theSet{0, \dots, n-1}$.

\mypara{Example}
Consider again \Cref{fig:runs-lookup}. The Figure depicts two runs
$\leftOf{\pi}$ and $\rightOf{\pi}$ of length 3 with respect to initial cycle~$1$
and source $\theSet{\var{in}}$ of the program in \Cref{fig:lookup}. Columns
$\var{in}$ and $\var{out}$ show $\mathit{store}(\pi,i)(in)$ and
$\mathit{store}(\pi,i)(out)$, for $\pi \in \theSet{\leftOf{\pi}$,
$\rightOf{\pi}}$ and $i \in \theSet{1,2}$. Similarly, columns
$\colorOf{\var{in}}$ and $\colorOf{\var{out}}$ show
$\mathit{live}(\pi,i)(\var{in})$ and $\mathit{live}(\pi,i)(\var{out})$, for $\pi
\in \theSet{\leftOf{\pi}$, $\rightOf{\pi}}$, and $i \in \theSet{1,2}$. The
Figure omits the initial state at cycle~$0$, where all liveness-bits are set to
$\deadCol$.

\mypara{Flushed, Constant-Time, Public}
For two runs $\leftOf{\pi}$ and $\rightOf{\pi}$ of length $n$, we say that
variable~$v$ is \emph{flushed}, if
$\mathit{store}(\leftOf{\pi},0)=\mathit{store}(\rightOf{\pi},0)$, we call $v$
\emph{public}, if
$\mathit{store}(\leftOf{\pi},i)=\mathit{store}(\rightOf{\pi},i)$, for $i \in
\theSet{0, \dots, n-1}$ and finally, we call $v$ \emph{constant-time}, if
$\mathit{live}(\leftOf{\pi},i)=\mathit{live}(\rightOf{\pi},i)$, for $i \in
\theSet{0, \dots, n-1}$.

\mypara{Secrecy Assumptions}
A set of secrecy assumptions~$\mathcal{A} \eqdef (\Flush, \Public)$ consists of
a set of variables~$\Flush \subseteq \textsc{Vars}$ that are assumed to be
flushed in the initial state, and a set of variables~$\Public \subseteq
\textsc{Vars}$, that are assumed equal throughout. A pair of runs $\leftOf{\pi}$
and $\rightOf{\pi}$ of length~$n$, satisfy a set of assumptions $\mathcal{A}$,
if, for each $v \in \Flush$, $v$ is flushed, and for each $v$ in \Public,  $v$
is public. We describe how \Tool synthesizes secrecy assumptions
in~\Cref{section:ctx-gen}.

\mypara{Constant-Time Execution}
We now define constant-time execution with respect to a set of
sinks~$\textsc{Snk} \subseteq \textsc{Vars}$, sources $\textsc{Src}$, and
assumptions $\mathcal{A}$.
We say that a program $P$ is constant-time, if for any initial cycle~$t$ and any
pair of runs $\leftOf{\pi}$ and $\rightOf{\pi}$ of $P$ with respect to $t$ and
$\textsc{Src}$ of length~$n$ that satisfy~$\mathcal{A}$, and any sink $o \in
\textsc{Snk}$, $o$ is constant-time.

\mypara{Example}
Consider again \Cref{fig:runs-lookup}. If we assume that variables in cycle~$0$
have the same value as in cycle~$1$, then $\var{out}$ is flushed
while~$\var{in}$ is not. Neither~$\var{in}$, nor $\var{out}$ are public, but
both are constant-time. As $\var{out}$ is constant-time in all runs, the program
in \Cref{fig:lookup} is constant-time with respect to the empty set of
assumptions and sink~$\theSet{out}$. In \Cref{fig:run}, none of the variables
are public or constant-time, however, the program in \Cref{fig:pipeline} can be
shown to be constant-time with $\Public = \theSet{\var{IF_pc}, \var{reset}}$.

\begin{figure}
  \begin{equation*}
    \begin{array}{@{}l@{\;\;}r@{}}
       \label{eq:horn}
       \mathit{init}(\leftOf{vs},\rightOf{vs}) \mathrel{\land} \mathit{flush} \mathrel{\land} \mathit{pub} \mathrel{\limp} \mathit{inv}(\leftOf{vs}, \rightOf{vs}, 0, t) & \mathit{(init)}\\[\jot]
        \\
       \begin{array}[c]{@{}c@{}}
        \mathit{inv}(\leftOf{vs}, \rightOf{vs}, c, t) \mathrel{\land}
        %
        \mathit{next}(\leftOf{vs},\rightOf{vs},\leftOf{vs}',\rightOf{vs}',t) 
      \mathrel{\land} \mathit{pub}
      \end{array}
      \limp \mathit{inv}(\leftOf{vs}', \rightOf{vs}', c+1, t)  & \mathit{(cons)}\\[\jot]
      \\
      \mathit{inv}(\leftOf{vs}, \rightOf{vs}, c, t) \mathrel{\land} \mathit{pub} \limp \leftOf{\colorOf{o}}=\rightOf{\colorOf{o}}, \; \mathit{for} \; o \in
      \textsc{Snk} & \mathit{(ct)}\\[\jot]
       
    \end{array}
  \end{equation*} 
  \caption{Horn clause encoding of the verification conditions for constant-time execution. \label{fig:vcs}}
  \vspace{-6ex}
\end{figure}

\subsection{Verifying Constant-Time Execution via Horn Constraints \label{sec:vcgen-cons}}
To verify constant-time execution of a circuit, we mirror the formal definition
of constant-time execution in a set of Horn clauses~\cite{Bjorner2015-sd}. We
start with the naive, monolithic encoding and discuss how to make it modular in
\Cref{sec:vcgen-sum}. At high level, the constraints \emph{i)}  issue a new live
instruction at a non-deterministically chosen initial cycle~$t$, and \emph{ii)}
ensure constant-time execution by verifying that the liveness-bits for each sink
are always the same, in any two runs.
The clauses---shown in \Cref{fig:vcs}---encode standard verification conditions
over an invariant $\mathit{inv}(\leftOf{vs}, \rightOf{vs}, c, t)$ of the product
circuit, where $vs$ ranges over all variables in the circuit and their respective
liveness bits, and $c$ and $t$ are the current and initial cycles, respectively.
We now describe the verification conditions in more detail.

\mypara{Initial States and Transition Relation}
Formula~$\mathit{init}(\leftOf{vs},\rightOf{vs})$ describes the product circuit's
initial states and requires all liveness-bits to be set to~$\deadCol$. To ensure
that the proof holds for any initial cycle, $\mathit{init}$ does not constrain
$t$.
Formula $\mathit{next}$ encodes the transition relation of the product circuit.
Like $\rightsquigarrow$, $\mathit{next}$ sets liveness-bits
of all sources to~$\activeCol$ at clock cycle~$t$.
Importantly, constructing~$\mathit{next}$ requires inlining all modules and
therefore can lead to large constraints that are beyond the abilities of the
solver.

\mypara{Assumptions}
For a set of assumptions~$\mathcal{A} \eqdef (\Flush,\Public)$, we construct
formulas~$\mathit{flush}$ and $\mathit{pub}$, both of which require the
variables in their respective sets to be equal in the two runs. We let
$\mathit{flush} \eqdef \left ( \wedge_{x \in \Flush}\ \leftOf{x}=\rightOf{x}
\right )$ and $\mathit{pub} \eqdef \left ( \wedge_{x \in \Public}\
\leftOf{x}=\rightOf{x} \right )$.
%

%

\mypara{Horn Constraints \& Solutions}
We then require that the invariant holds,initially ($\mathit{init}$), assuming
all variables in~$\Flush$ and $\Public$ are equal in both runs; that the
invariant is preserved under the transition relation of the product circuit,
assuming that public variables are equal in both runs ($\mathit{cons}$), and
finally, that the liveness-bits of any sink are the same in both runs
($\mathit{ct}$).
These constraints are then passed to an off-the-shelf
solver~\cite{liquid-fixpoint} yielding a formula which, when substituted
for~$\mathit{inv}$, makes all implications valid and thus proves constant-time
execution.

\mypara{Proof Artifacts}
While attempting to construct the invariant, the solver's internal
representation keeps track of \emph{i)} the set of variables which it was able
to prove to be constant-time and public and \emph{ii)} the \emph{order} in which
the remaining variables lost the respective properties. In case of a failed
proof attempt, the prover communicates these artifacts to the assumption
synthesis part of \Tool, which in turn uses them to identify the verification
failure, as we will discuss in the next section.  

\subsection{Finding Modular Invariants \label{sec:vcgen-sum}}
Naively, constructing~$\mathit{next}$ requires all the code to be in a
\emph{single} module. However, this can yield gigantic circuits whose Horn
clauses are too large to analyze efficiently.
To avoid instantiating the entire module at each usage site, \Tool constructs
\emph{module summaries} which concisely describe the timing relevant properties
of the module's input and output ports. 

\mypara{Per-Module Invariants and Summaries}
Instead of a single whole program invariant $\mathit{inv}$, the modular analysis
constructs a per-module invariant~$\mathit{inv}_m$, and an additional
summary~$\mathit{sum}_m$, for each module~$m$. The summary only ranges over
module inputs and outputs, and respective liveness-bits ($\var{io}$) and needs
to be implied by the invariant, \ie we add a clause stating that
$\mathit{inv}_m(\leftOf{vs},\rightOf{vs},t) \Rightarrow
\mathit{sum}_m(\leftOf{io},\rightOf{io},t)$.
The analysis produces the same constraints as before, but now on a per-module
basis, that is, we require module invariants to hold on initial states
($\mathit{init}$), and be preserved under the transition relation
($\mathit{cons}$), but, instead of using the overall transition
relation~$\mathit{next}$ we use a per-module transition
relation~$\mathit{next}_m$. It may now happen that $\mathit{next}_m$ makes use
of a module~$n$, but instead of inlining the transition relation of $n$ as
before, we substitute it by its module summary~$\mathit{sum}_n$, thereby
avoiding the blowup in constraint size. Finally, we restrict sources and sinks
to occur at the top-level module, and add a clause requiring that any sink has
the same liveness-bits in both runs ($\mathit{ct}$). The summaries are also used to
modularize our assumption synthesis algorithm~\cref{sec:cex-modules}.

\section{Counterexamples \& Assumption Synthesis}
\label{section:ctx-gen}
We now explain how \Tool uses the proof artifacts to help the user
understand and explicate secrecy assumptions when verification fails.
We first describe how \Tool analyzes the artifacts from the failed proof attempt
in order to compute a \emph{counterexample} consisting of the set of variables
that---according to the information communicated by the prover---lost the
constant-time property first (\Cref{sec:min-cex}). Next, we discuss how \Tool
uses the counterexample to synthesize a set of \emph{secrecy assumptions} that
eliminate the root cause of the verification failure (\Cref{sec:cex-assum}).
This is done by computing a \emph{blame-set} which contains the set of variables
that likely caused the loss of constant-time for the variables in the
counterexample via a control dependency. This blame set is then used to encode
an optimization problem whose solution determines a minimal set of assumptions
required to remove the timing violation.
Finally, we briefly discuss how \Tool uses module summaries to speed up
counterexample generation and secrecy assumption synthesis
(\Cref{sec:cex-modules}).

\subsection{Computing Counterexamples \label{sec:min-cex}}

\mypara{Dependency Graph}
To compute the counterexample from a failed proof attempt, \Tool first creates a
\emph{dependency graph} $G \eqdef (V, D \cup C)$ which encodes data-, and
control dependencies between program variables.
$G$ consists of \emph{variables}~$V \subseteq \textsc{Vars}$,
\emph{data-dependencies}~$D \subseteq (\textsc{Vars} \times \textsc{Vars})$,
where~$(v,w) \in D$ if $v$'s value is used to compute~$w$ directly through an
assignment, and \emph{control-dependencies}~$C \subseteq (\textsc{Vars} \times
\textsc{Vars})$ where $(v,w) \in C$, if ~$v$'s value is used indirectly, \ie
$w$'s value is computed under a branch whose condition depends on~$v$. 

\mypara{Variable-Time Map}
Next, \Tool extracts an artifact from the failed proof attempt: a partial map
$\nonctMap \in (\textsc{Vars} \rightharpoonup \mathbb{N})$ which records the
\emph{temporal order} in which variables started to exhibit timing variability.
Importantly, if for some $v$, $\nonctMap$ is undefined (we write
$\nonctMap(v)=\bot$), the solver was able to prove that $v$ is constant-time.
For any other variables $v$ and $w$, if $\nonctMap(v) < \nonctMap(w)$, then $v$
started to exhibit timing variability \emph{before} $w$.\footnote{More formally,
for variables $v,w$ if $\nonctMap(v) < \nonctMap(w)$, then there exist two
runs~$\leftOf{\pi}$ and $\rightOf{\pi}$ of some length~$n$, and two numbers $0
\leq i,j < n$ such that $i$ is the smallest number such that
$\mathit{live}(\leftOf{\pi},i)(v) \neq \mathit{live}(\rightOf{\pi},i)(v)$ and
similarly $j$ is the smallest number such that $\mathit{live}(\leftOf{\pi},j)(w)
\neq \mathit{live}(\rightOf{\pi},j)(w)$ and $i < j$. This information can \eg be
extracted from a concrete counterexample trace, like the one shown
in~\Cref{fig:run}. }
\Tool uses this map to break cyclic data-flow dependencies.

\mypara{Computing the Counterexample}
Using the data-flow graph, and $\nonctMap$, \Tool computes a \emph{reduced
graph}.
\Tool removes from the dependency graph, all nodes that are constant-time and
all edges $(w,v)$ such that~$\nonctMap(w) > \nonctMap(v)$.
Intuitively, if variable $w$ has started to exhibit timing variability
\emph{after} variable~$v$, it cannot be the cause for $v$ losing the
constant-time property.
Finally, \Tool removes all nodes that cannot reach a sink node using the
remaining edges.
This leaves us with a set of variables~$\CEX \subseteq \textsc{Vars}$ without
incoming edges, which we identify as the root cause of the violation, and which
we present---as counterexample---to the user. We now define the reduced graph in
more detail.

\mypara{Reachability}
For dependency graph $G \eqdef (V, D \cup C)$ and nodes $v,w \in V$ we write~$v
\rightarrow w$, if $(v,w) \in (D \cup C)$, $v \rightarrow^n w$, if there is a
sequence $v_0 v_1 \dots v_{n-1}$, such that $v_0=v$ and $v_{n-1}=w$, and $v_i
\rightarrow v_{i+1}$ for $i \in \theSet{0, \dots, n-2}$. Finally, we say $w$ is
reachable from $v$, if there exists $n$ such that $v \rightarrow^n w$.

\mypara{Reduced Graph}
For a data-flow graph~$G \eqdef (V,D \cup C)$, and map $\nonctMap$, we define
the reduced graph with respect to $\nonctMap$  as the
largest subgraph $G' \eqdef (V',D' \cup C')$ such that $V' \subseteq V$, $D'
\subseteq D$, $C' \subseteq C$ and
\begin{enumerate}
\item No node is constant-time, \ie for all $v \in V'$, $\nonctMap(v) \neq
\bot$.
\item All edges respect the causal order given by~$\mathit{vartime}$, \ie for
  all $(v,w) \in (D' \cup C')$, we have $\nonctMap(v) \leq \nonctMap(w)$.
\item All nodes can reach a sink, \ie for all $v \in V'$, there is $o \in
  \textsc{Snk}$ such that $o$ is reachable from~$v$.
\end{enumerate}
For variable~$v$, and graph~$G \eqdef (V, D \cup C)$, let $\mathit{pre}(v,G)$ be
the set of its immediate predecessors in $G$, that is $\mathit{pre}(v,G) \eqdef
\theSet{w \; | \; (w,v) \in (D \cup C)}$.
We define the counterexample~$\CEX$ of a graph $G$ with map
$\nonctMap$ as the set of nodes in the reduced graph~$G'$ (wrt. $\nonctMap$),
that have no predecessors, \ie $\CEX \eqdef \theSet{v \ | \
\mathit{pre}(v,G') = \emptyset}$.

\mypara{Example: Simplified Pipeline}
The code in \Cref{fig:pipeline-simple} shows a simplified version of the
pipelined processor from \Cref{fig:pipeline}.
\begin{wrapfigure}[18]{l}{.62\textwidth}
  \footnotesize	
  \centering
  \resizebox{.7\linewidth}{!}{
\begin{lstlisting}[style={verilog-style}]
assign ID_rt = ID_instr[20:16]; (* \label{code:ex1-set-target} *)

rom32 IMEM(IF_pc, IF_instr); (* \label{code:ex1-inst-fetch} *)

always @((***))
  stall = (ID_rt == EX_rt) (* \label{code:ex1-set-stall} *)

always @(posedge clk) begin
  if (Stall == 1) begin (* \label{code:ex1-if} *)
      ID_instr  <= ID_instr; (* \label{code:if-stall} *)
      EX_rt     <= EX_rt;
  end else begin
      ID_instr  <= IF_instr; (* \label{code:if-no-stall} *)
      EX_rt     <= ID_rt;
  end
end
\end{lstlisting}
  }
  \caption{MIPS Pipeline Fragment in Verilog.}
  \label{fig:pipeline-simple}
\end{wrapfigure}
Like in \Cref{fig:pipeline}, the pipeline either stalls (\Cref{code:if-stall})
if flag \var{Stall} is set (\Cref{code:ex1-if}), or else forwards values to the next
stage (\Cref{code:if-no-stall}).
To avoid a write-after-write data-hazard, the \var{Stall} flag is set, if the
instructions in the execute and decode stage have the same target registers
(\Cref{code:ex1-set-stall}). The target register is calculated from the current
instruction~(\Cref{code:ex1-set-target}), and the instruction is, in turn,
fetched from memory using the current program counter~(\Cref{code:ex1-inst-fetch}).
Note the cyclic dependency between \var{ID_instr} and \var{Stall}
that turns comprehending the root cause into a ``chicken-and-egg''
problem.

\mypara{Dependency Graph}
To check if the pipeline fragment executes in constant-time, we mark
\var{IF_pc} as source, and \var{ID_instr} as sink and run \Tool.
Since the pipeline is variable-time, the verification fails.
To compute a minimal counterexample, \Tool creates the dependency graph shown in
\Cref{fig:ex-1-var-dep}.
Each node is annotated with information extracted from the failed
proof attempt: the node is labeled with its value under~$\nonctMap$,
and is marked with (\textcolor{green}{\checkmark}) if \Tool was able
to verify that the variable is constant-time and (\textcolor{BrickRed}{\text{\sffamily X}})
otherwise.
Solid edges represent data-, and dashed edges represent control dependencies.
The ordering induced by $\nonctMap$ allows us to break the cyclic dependency
between variables $\var{Stall}$ and $\var{ID_instr}$, thereby resolving the
chicken-and-egg problem.

\begin{figure*}[t!]
  \centering
  \subfloat[Dependency Graph.]{
  \begin{minipage}[c]{.49\textwidth}
  \resizebox{\linewidth}{!}{
    \begin{tikzpicture}[>=latex,line join=bevel,node distance=0.4cm and 0.6cm]

      \node (n1) [thick,draw,ellipse] {IF_pc (\greencheck, $\bot$)};
      \node (n2) [thick,draw,ellipse, above= of n1] {IF_inst (\greencheck, $\bot$)};
      \node (n3) [thick,draw,ellipse, right = of n2] {ID_instr (\redcross, $1$)};
      \node (n4) [thick,draw,ellipse, above right = of n3] {Stall (\redcross, $3$)};
      \node (n5) [thick,draw,ellipse, below right= of n3] {ID_rt (\redcross, $2$)};
      \node (n6) [thick,draw,ellipse, below right = of n4] {EX_rt (\redcross, $3$)};

      \draw [->,solid] (n1) edge (n2);
      \draw [->,solid] (n2) edge (n3);
      \draw [->,solid] (n3) edge (n5);
      \draw [->,solid] (n5) edge (n4);
      \draw [->,solid] (n5) edge  (n6);
      \draw [->,solid] (n6) edge [bend left] (n4);
      \draw [->,dashed] (n4) edge [bend right](n3);
      \draw [->,dashed] (n4) edge [bend left] (n6);
    ;
      \end{tikzpicture}
    }
    \label{fig:ex-1-var-dep}
  \end{minipage}
  }
  \quad
  \subfloat[Dependency Graph after eliminating constant-time nodes and edges
  violating the order given by $\nonctMap$.]{
    \begin{minipage}[c]{.35\linewidth}
        \resizebox{\linewidth}{!}{
    \begin{tikzpicture}[>=latex,line join=bevel,node distance=0.4cm and 0.6cm]
      \node (n3) [thick,draw,ellipse, right = of n2] {ID_instr (\redcross, $1$)};
      \node (n4) [thick,draw,ellipse, above right = of n3] {Stall (\redcross, $3$)};
      \node (n5) [thick,draw,ellipse, below right= of n3] {ID_rt (\redcross, $2$)};
      \node (n6) [thick,draw,ellipse, below right = of n4] {EX_rt (\redcross, $3$)};

      \draw [->,solid] (n3) edge (n5);
      \draw [->,solid] (n5) edge (n4);
      \draw [->,solid] (n5) edge  (n6);
      \draw [->,solid] (n6) edge [bend left] (n4);
      \draw [->,dashed] (n4) edge [bend left] (n6);
    ;
      \end{tikzpicture}
    }
  \label{fig:ex-1-root-causes}  
  \end{minipage}
  }

  \subfloat[Dependency Graph with Module Summary.]{
   \begin{minipage}[c]{.55\textwidth}
    \resizebox{\linewidth}{!}{
    \begin{tikzpicture}[>=latex,line join=bevel,node distance=0.4cm and 0.6cm]

      \node (n1) [thick,draw,ellipse] {IF_pc (\greencheck, $\bot$)};
      \node (n2) [thick,draw,ellipse, above= of n1] {IF_inst (\greencheck, $\bot$)};
      \node (n2a) [dashed,draw,ellipse, above= 1.2cm of n2] {IF_instr (\greencheck, $\bot$)};
      \node (n3) [thick,draw,ellipse, right = of n2] {ID_instr (\redcross, $1$)};
      \node (n3a) [dashed,draw,ellipse, above= 1.2 cm of n3] {ID_instr (\redcross, $1$)};
      \node (n4) [thick,draw,ellipse, above right = of n3] {Stall (\redcross, $3$)};
      \node (n4a) [dashed,draw,ellipse, above = 1cm of n4] {Stall (\redcross, $3$)};
      \node (n5) [thick,draw,ellipse, below right= of n3] {ID_rt (\redcross, $2$)};
      \node (n6) [thick,draw,ellipse, below right= of n4] {EX_rt (\redcross, $3$)};

       \draw [->,solid] (n1) edge (n2);
       \draw [->,solid] (n2) edge (n2a);
       \draw [->,solid] (n2a) edge (n3a);
       \draw [->,solid] (n3a) edge (n3);
       \draw [->,solid] (n4) edge (n4a);
       \draw [->,solid] (n3) edge (n5);
       \draw [->,solid] (n5) edge (n4);
       \draw [->,solid] (n5) edge  (n6);
       \draw [->,solid] (n6) edge [bend left] (n4);
       \draw [->,dashed] (n4a) edge [bend right](n3a);
       \draw [->,dashed] (n4) edge [bend left] (n6);
       \node [label={\text{Summary Graph}}] (box) [draw,rounded corners, dashed, fit = (n2a) (n3a) (n4a)] {};

       \label{fig:ex-1-var-dep-summary}
     \end{tikzpicture}
   }
 \end{minipage}
 }
    \caption{\Cref{fig:ex-1-var-dep} shows the dependency graph for \Cref{fig:pipeline-simple}.
      Data-dependencies are shown as solid edges, and control-dependencies are
      shown dashed.
      Each node is labeled with its $\nonctMap$-value and marked
      (\textcolor{green}{\checkmark}) if \Tool was able to prove
      the variable constant-time and (\textcolor{BrickRed}{\text{\sffamily X}})
      otherwise.
      \Cref{fig:ex-1-root-causes} shows the dependency graph after eliminating constant-time nodes from \Cref{fig:ex-1-var-dep},
      and removing edges that violate the variable-time map. Removing the edge between
      \var{Stall} and \var{ID_instr} breaks the cyclic dependency in the
      original graph.
      %
      %
      Figure \ref{fig:ex-1-var-dep-summary} shows the variable dependency graph
      with a summary-graph extracted from the module summary.
      %
      }
\end{figure*}

\mypara{Reduced Dependency Graph}
\Cref{fig:ex-1-root-causes} shows the dependency graph after removing all
constant-time nodes and edges that violate the causal ordering.
\Tool erases all nodes that cannot reach
sink~\var{ID_instr}. This only leaves
\var{ID_instr} which we return as counterexample.

\mypara{Remark}
In case the proof artifact only partially resolves the cyclic dependencies, that
is, $\mathit{vartime}$ only defines a partial order over non-constant-time
variables, the reduced graph may still contain cycles, and therefore there may
be no nodes without predecessor. We can however still apply our technique by
computing the graph's strongly connected components, and including all nodes in
the respective component in the counterexample.

\subsection{Assumption Synthesis \label{sec:cex-assum}}
The previous step leaves us with a set of nodes~\CEX, which lost the
constant-time property first. 
Since these nodes must have lost the constant-time property through a control
dependency on a secret value, we can compute a set of variables \textsc{Blame}
that are directly responsible: the immediate predecessors of \CEX in
the dependency graph with respect to a control dependency.
Formally, for dependency graph~$G=(V,D \cup C)$, we
let $\textsc{Blame} \eqdef \theSet{w \; | \; v \in \CEX \land (w,v) \in
C}$.
To synthesize secrecy assumptions that remove the constant-time violation, we
could directly assume that all nodes in \textsc{Blame} are public.
But this is often a poor choice: variables in \textsc{Blame}
can be defined deep inside the circuit, whereas we would like to phrase
our assumptions in terms of externally visible \emph{input sources}.

\mypara{Finding Secrecy Assumptions via ILP}
Instead, we compute a minimal set of assumptions close to the input sources
via a reduction to Integer Linear Programming (ILP).
To this end, we use a second proof artifact, a map $\secretMap$ that---similar
to $\nonctMap$---describes the temporal order in which the verifier determines
variables have become \emph{secret}, \ie ceased being public.
Let $G'=(V',D' \cup C')$ be the reduced dependency graph with
respect to $\secretMap$, and let $\textsc{No} \subseteq V'$ be
a set of variables that the user chose to exclude from consideration.
\Tool produces constraints on a new set of variables: two constraint variables
$m_v \in \{ 0, 1 \}$ and $p_v \in \{ 0, 1 \}$, for each program variable~$v$,
such that $m_v=1$, if program variable $v$ is \emph{marked} public by an
assumption, and $p_v=1$, if $v$ can be \emph{shown to be} public, that is, it is
either marked public, or all its predecessors are public.
Then, \Tool produces the following set of constraints.
\begin{equation*}
  \begin{array}[c]{@{}l@{\quad}l@{\quad}l@{}}
    \ilpMarked{v} \geq \ilpPublic{v}, & \mathit{if} \ v \in V' \ , \ \mathit{pre}(v,G') = \emptyset & (1) \\[2ex]
    \ilpMarked{v} + \left ( \frac{\sum_{w \in \mathit{pre}(v,G')} \ilpPublic{w}}{\#\mathit{pre}(v,G')} \right )  \geq \ilpPublic{v} & \mathit{if} \ v \in V' \ , \ \mathit{pre}(v,G') \neq \emptyset & (2) \\[2ex]
    \ilpPublic{v}=1 & \mathit{if} \ v \in \left ( \textsc{Blame} \setminus \textsc{No} \right )  & (3) \\[2ex]
    \ilpMarked{v}=0 & \mathit{if} \ v \in \textsc{No} & (4) 
  \end{array}
\end{equation*}
Constraints $(1)$ and $(2)$ ensure that a variable is public, if either it is marked
public, or all its predecessors in $G'$ are public. Constraint $(3)$ ensures
that all blamed variables that have not been excluded can be shown to be public,
and finally, constraint $(4)$ ensures that all excluded constraints are not
marked. Let $d(v,w)$ be a distance metric, \ie a function that maps pairs of
nodes to the natural numbers. Then we want to solve the constraints using the
following objective function that we wish to minimize, where for $v \in V'$, we
define as
weight the minimal distance from one of the source nodes~$w_v = \left ( \argmin_{in \in \textsc{Src}} d(in, v) \right )$:
\begin{equation}
  \label{eq:ilp-obj}
  \tag{objective}
  \sum_{v \in V'} w_v \ilpMarked{v}.
\end{equation}

\noindent
A solution to the constraints defines a set of
assumptions~$\mathcal{A}=(\textsc{Flush}, \textsc{Pub})$, where we
let~$\textsc{Flush} \eqdef \theSet{v \in V' \ | \ \ilpMarked{v}=0,
\ilpPublic{v}=0 }$ and $\textsc{Pub} \eqdef \theSet{v \in V' \ | \
\ilpMarked{v}=1}$. The constraints can be solved efficiently by an off-the-shelf
ILP solver.

\mypara{Example: Simplified Pipeline}
Consider again the simplified pipeline in \Cref{fig:pipeline-simple}.
As we identified \var{ID_instr} as the root cause in the
previous step, we need to ensure that its blame-set consisting of all indirect
influences is public.
\var{ID_instr} only depends on \var{Stall}, and
therefore we add constraint $\ilpPublic{Stall} = 1$. 
Since, all variables are secret (\ie we didn’t make any public-assumptions yet),
the reduced graph is equal to the original graph.
For variables~\var{IF_instr} and \var{ID_instr}, we get:
$\ilpMarked{IF_instr} + \ilpPublic{IF_pc} \geq \ilpPublic{IF_instr}$ and
$\ilpMarked{ID_Instr} + \frac{\ilpPublic{IF_instr} + \ilpPublic{Stall}}{2}  \geq \ilpPublic{ID_Instr}$.
We obtain the following objective function:
%
    $\ilpMarked{IF_pc} \; + \; 2 \ilpMarked{IF_instr} \; + \; 3 \ilpMarked{ID_instr} \; + \dots$.
%
Sending the constraints to an ILP solver produces a solution, where
~$\ilpMarked{IF_pc}=1$, $\ilpMarked{v}=0$, for all variables $v \neq
\var{IF\_PC}$, and $\ilpPublic{v}=1$, for all~$v$.
This corresponds to the following assumption set~$\mathcal{A} \eqdef (Flush,
\theSet{\var{IF\_PC}})$, where $Flush$ includes all variables except
\var{IF\_PC}.
This corresponds exactly to the desired minimal solution where we only mark
\var{IF_pc} as public.
Note that our method does not necessarily result in all variables becoming
public. We give an example in \Cref{sec:non-pub}.

\subsection{Modular Assumption Synthesis \label{sec:cex-modules}}
To avoid a blowup in constraint size, we want to avoid inlining instantiated
modules.
We therefore extract a dependency graph from the module summary:
whenever the summary requires an input~\var{in} to be public for an
output~\var{out} to be constant-time, we draw a control dependency between
\var{in} and \var{out}. Whenever the summary requires an input \var{in} needs to be
constant-time for an output \var{out} to be constant-time, we draw a data dependency.
Finally, we insert the computed summary graph into the top level dependency
graph, and connect the instantiation parameters to the graph's inputs
and outputs.

\mypara{Example}
We modify \Cref{fig:pipeline-simple} to factor out the updates to
\var{ID_instr} into a separate module. \Tool computes the following summary
invariant, from which we create the graph in~\Cref{fig:ex-1-var-dep-summary}:
%
    $\ctVar{IF\_instr} \mathrel{\land}  \pubVar{Stall} \Rightarrow \pubVar{ID\_instr}$.
%
Since connecting the instantiated variables to the summary graph is equivalent
to the original graph (\Cref{fig:ex-1-var-dep}), our analysis returns the same
result.

\section{Implementation}

\sys is split into front-end and back-end.
Our front-end translates \Verilog to the \Iodine intermediate representation
(IR)~\cite{iodine} and associates secrecy assumptions with
input and output wires.
Our back-end translates this annotated IR into verification
conditions (Horn clauses); when verification fails, we generate counter
examples and secrecy assumptions and present them to the user for feedback.
We implement the back-end in roughly 9KLOC Haskell, using the
\texttt{liquid-fixpoint} (0.8.0.2)~\cite{liquid-fixpoint} and \textsc{Z3}
(4.8.1)~\cite{Z3} libraries for verification, and the \textsc{GLPK}
(4.65)~\cite{glpk} library for synthesizing assumptions by solving the ILP
problem of \Cref{section:ctx-gen}.
Our tool and evaluation data sets (described next) are open source and available
on GitHub.
\footnote{We omit the link for the double-blind review process.}

\section{Evaluation}
\label{sec:evaluation}

\begin{table*}[t]
  \centering
  \footnotesize
  \resizebox{.95\linewidth}{!}{%
  \begin{tabular}{lrrrcrrrrrr}
    \toprule

    \multirow{2}{*}{\textbf{Name}}
    & \multirow{2}{*}{\textbf{\#LOC}}
    & \multicolumn{2}{c}{\textbf{\#Assum}}
    & \multirow{2}{*}{\textbf{CT}}
    & \multicolumn{2}{c}{\textbf{Check (H:M:S)}}
    & \multirow{2}{*}{\textbf{\# Iter}}
    & \multirowcell{2}{\textbf{CEX} \\ \textbf{Ratio}}
    & \multirowcell{2}{\textbf{Sugg} \\ \textbf{Ratio}}
    & \multirowcell{2}{\textbf{Accept} \\ \textbf{Ratio}} \\

    & & \textbf{\#flush} & \textbf{\#public} & & \textbf{Inlined} & \textbf{Modular} \\

    \midrule

    MIPS~\cite{MIPS}
    & 447 & 28 & 3 & \tickYes & 2.42 & 3.13
    & 3 &  2.50\%  & 1.73\% & 83.33\% \\

    RISC-V~\cite{Yarvi}
    &  514 &  10 &  11 & \tickYes & 13.21 & 10.23
    &  5 & 16.24\%  & 3.98\% &  46.90\% \\

    SHA-256~\cite{Sha-core}
    &  563 &  4 &  3 & \tickYes & 7.21 & 8.90
    &  2 &  4.28\%  & 3.57\% & 100.00\% \\

    FPU~\cite{FPU}
    & 1108 &  3 &  1 & \tickYes &  9.10 & 11.54
    &  1 &  0.33\%  & 0.26\% & 100.00\% \\

    ALU~\cite{CTALU}
    & 1327 &   1 &  3 & \tickYes &  2.01 &  2.29
    &  2 &  0.88\%  & 1.38\% &  75.00\% \\

    FPU2~\cite{FPU2}
    &  272 &  24 &   4 &  \tickNo &  1.31 &  3.65
    &  - &  -  & - &  - \\

    RSA~\cite{RSA}
    &  855 &  29 & 4 &  \tickNo &  2.87 &  1.51
    &  - &  - & - &  - \\

    \midrule

    AES-256~\cite{tiny-aes}
    &  800 &   0 &   0 & \tickYes & 6:05:01.82 &    2.74
    &  - &  - & - &  -  \\

    SCARV~\cite{scarv-cpu}
    & 8468 &  73 &  54 & \tickYes &   14:20.93 & 8:35.46
    & 34 &  9.08\%  & 5.68\% &  84.80\% \\

    \midrule
    \midrule

    \textbf{Total}
    & 14354 & 159 & 89 & - & 6:20:00.88 & 9:19.45
    & 47 & 5.55\%* & 2.77\%* & 81.67\%* \\

    \bottomrule
  \end{tabular}%
  }
  \caption{
\textbf{\#LOC} is the number of lines of Verilog code (without comments or empty
lines),
\textbf{\#Assum} is the number of assumptions; \textbf{flush} and
\textbf{public} are sizes of the sets $\Flush$ and $\Public$ respectively,
\textbf{CT} shows if the program is constant-time,
\textbf{Check} is the time \Tool took to check the program; \textbf{Inlined}
and \textbf{Modular} represent inlining module instances and using module
summaries respectively.
\textbf{\# Iter} is the number of times the user has to invoke \Tool to verify
the benchmark starting with an empty set of assumptions;
\textbf{CEX Ratio} is the average ratio of the number of identifiers in the
counterexample to all variable-time identifiers in a given iteration;
\textbf{Sugg Ratio} is the average ratio of the number of secrecy
assumptions that \Tool suggests to all secret variables in a given
iteration,
and \textbf{Accept Ratio} is the average ratio of the suggested assumptions
accepted by the user.
In the \textbf{Total} row, we use * to denote averages instead of sums. 
We do not run error localization on FPU2 and RSA because they
are variable-time; AES-256 does not need any assumptions.
%
%
}
\label{tab:evaluation}
\end{table*}

We evaluate \Tool by asking the following questions:
\begin{CompactItemize}
\item\textbf{Q1:} Are constant-time counterexamples effective at localizing the
cause of verification failures?
\item\textbf{Q2:} Are the secrecy assumptions suggested by \Tool useful?
\item\textbf{Q3:} What is the combined effect of counterexamples and
secrecy assumption generation on the verification effort?
\item\textbf{Q4:} Do module summaries improve scalability?
\end{CompactItemize}

\noindent
To answer questions \textbf{Q1} and \textbf{Q2}, we use \Tool to recover the
assumptions for the benchmark suite from~\cite{iodine}. These benchmarks include
a MIPS and RISC-V core, ALU and FPU modules, and RSA and SHA-256 crypto modules.
To evaluate questions \textbf{Q3} and \textbf{Q4},  we evaluate \Tool on two
challenging new benchmarks, the SCARV ``side-channel hardened RISC-V'' processor
\cite{scarv-cpu} whose size exceeds the largest benchmark from~\cite{iodine} by a
factor of 10, and a highly modular AES-256 implementation~\cite{tiny-aes}.

\mypara{Summary}
\Tool's counterexample synthesis dramatically reduces the number of
potential error locations users have to manually inspect (6\% of its original
size) and most of \Tool's assumption suggestions are accepted by the user
(on average 81.67\%).
Module summaries are key to reducing verification times for certain hardware
designs (\eg for AES-256 crypto core summaries reduced the verification time
from six hours to a three seconds).
We find the counterexamples and secrecy assumptions suggested by \Tool
to be crucial to reducing the human-in-the-loop time from days to (at worst)
hours.

\mypara{Experimental Setup}
We run all experiments on a 1.9GHz Intel Core i7-8650U machine with 16~GB of
RAM, running Ubuntu 20.04 with Linux kernel 5.4.

\mypara{Methodology}
For every benchmark, we start with an empty set of secrecy assumptions and run
\Tool repeatedly to recover the missing assumptions needed to verify the benchmark.
We collect the following information after every invocation of the tool:
the total number of variables that are variable-time and secret; the size of the
counterexamples measured by the number of variables they contain; the number of
assumptions \Tool suggests, and the number of assumptions we reject during each
iteration; finally, we record the number of times we invoke \Tool to complete each
verification task.
With all the assumptions in place, we measure the time it takes for the tool to
verify each benchmark; we report the median of thirty runs for all but the
non-modular (inlined) AES benchmark, for which---due to its size---we report the
median of three runs.

\mypara{Q1: Error Localization} To understand whether our counterexample
generation is effective at localizing the cause of verification failures, we
compare the number of variables in the counterexample to the total number of
non-constant-time variables.
The \textbf{CEX Ratio} column \Cref{tab:evaluation} reports the average ratio
per iteration.
On average, we observe that less than 6\% of non-constant-time variables are
included in the counterexample.
Since the total number of non-constant-time variables is typically on the order
of hundreds (\eg the median (and geomean) number of  non-constant-time variables
across all benchmarks and iteration is 97 (94)), this dramatically reduces the
number of variables the developer has to inspect in order to understand the
violation.
For the benchmarks that were variable time, the counterexamples also precisely
pinpointed where in the circuit the constant-time property was violated.
For example, in the FPU2 benchmark \Tool included the \var{state} register in
its third iteration counterexample.
This register indicates when the FPU's output is ready.
Inspecting the register's blame-set (similar to the process described
in~\Cref{sec:overview-secrecy}) revealed that its value is set depending on
whether one of the operands to the division operation is NaN and thus the FPU
clearly leaks information about its operands.

\mypara{Q2: Identifying Secrecy Assumptions}
To assess the quality of secrecy assumptions suggested by \Tool, we record the
number of suggestions that the user accepts (\emph{useful} suggestions) and the
ratio of suggestions to the total number of secret variables the user would
otherwise have to inspect manually.
We find that most (on average 81.67\%) of \Tool's suggestions are useful,
reported in the \textbf{Accept Ratio} column of \Cref{tab:evaluation}.
Moreover, we observe that the number of variables included in the
counterexamples is relatively small (\textbf{Sugg Ratio} column); on average, we
only had to inspect 2.77\% of the secret variables.

\mypara{Q3: Verification Effort}
Finally, as a rough measure of the overall verification effort, we count the
number of user interactions, \ie the number of times we invoked \Tool after
modifying our set of secrecy assumptions.
Verifying the largest benchmark from~\cite{iodine}, the \textsc{Yarvi} RISC-V
core~\cite{Yarvi} took five invocations over several minutes.
The final assumptions we arrived at were the same as the assumptions manually
identified by the authors of \Iodine in~\cite{iodine}; they, however, took
multiple days to identify these assumptions and verify this
core~\cite{personal-rami}.
Verifying the SCARV core took thirty four iterations and roughly three hours;
this core is considerably larger (roughly 10$\times$) than the \textsc{Yarvi} RISC-V core 
and, we think, beyond what would possible with tools
like \Iodine, which rely on manual annotations and error localization.
Indeed, we found the error localization and assumption inference to be
especially useful in narrowing our focus and understanding to small parts of the
core and avoid the need to understand complex implementation details irrelevant
to the analysis.
%

\mypara{Q4: Scalability}
To evaluate how module summaries affect the scalability, we compare the time it
takes to verify (or show variable-time) a program with and without module
summaries.
Columns \textbf{Inlined} and \textbf{Modular} of \Cref{tab:evaluation} gives
the run times of \Tool with inlining (no summaries) and module summaries,
respectively.
On the \Iodine benchmarks (the first seven benchmarks), we observe that module
summaries don't meaningfully speed up verification.
Indeed, on average, module summaries only reduce the size of the query sent to
our solver by roughly 5\% on these benchmarks.
On the more complex AES-256 and SCARV benchmarks, however, we see the benefit
of module summaries.
For AES-256, using module summaries reduces the query size by 99.7\%, from
391.3MB to 1.2MB, which, in turn, reduces the verification time by three orders
of magnitude---from six hours to three seconds.
Module summaries allow \Tool to exploit the core's modular design, i.e.,
AES-256's multiple and nested instantiations of the same modules (see
\Cref{fig:aes-module-dep}).
For SCARV, summaries reduce the query size by 41\%, and speed up the
verification time by 40\%.
Though this reduction is not as dramatic as the AES-256 case, the speedup did
improve \Tool's interactivity.
\section{Related Work}
\label{sec:related-work}

\mypara{Verifying Leakage Freedom}
There are various techniques, such as \texttt{ct-verif}~\cite{CTSoftware},
\cite{barthe:ct}, and \textsc{CT-Wasm}~\cite{watt:2019:ctwasm}, that verify
constant-time execution of software, and quantify leakage through timing and
cache side-channels~\cite{MyersPLDI2012, Koepf2013, rodrigues2016sparse,
almeida2013formal, tis-ct, ctgrind}.
However, their analyses do not directly apply to our setting:
They consider straight-line, sequential code, unlike the highly parallel nature
of hardware.
%
%
There are many techniques for verifying \emph{information flow} properties of hardware.
Kwon et al.~\cite{Harris17} prove information flow safety of hardware for
policies that allow explicit declassification and are expressed over streams of
input data.
SecVerilog~\cite{secverilog} and Caisson~\cite{li2011caisson} use information
flow types to ensure that generated circuits are secure.
GLIFT~\cite{tiwari2009complete, Kasnter11} tracks the flow of information at the
gate level to eliminate timing channels.
Other techniques such as HyperFlow~\cite{ferraiuolo2018hyperflow},
GhostRider~\cite{liu2015ghostrider} and Zhang et al.~\cite{MyersPLDI2012} take
the hardware and software co-design approach to obtain end-to-end guarantees.
\texttt{dudect}~\cite{dudect}, detects end-to-end timing variabilities across
the stack via a black-box technique based on statistical measurements.
\Iodine~\cite{iodine}, like \Tool, focuses on clock-precise constant-time execution,
not information flow.
Unlike \Tool, none of these methods provides help in elucidating secrecy
assumptions, in case the verification fails---a feature we found essential in
scaling our analysis to larger benchmarks. 
We see the techniques presented in this paper as complementary and would like to
explore their potential for scaling existing verification methods for hardware
and software.

\mypara{Fault Localization}
There are several approaches to help developers localize the
root causes of software bugs~\cite{fault-loc-survey}.
Logic-based fault localization techniques
\cite{Flow-Sensitive-Fault-Localization,ermis-error-invariants,joseBugAssistAssistingFault2011a,cause-clue-clauses}
are the closest line of work to ours.
For example, BugAssist~\cite{joseBugAssistAssistingFault2011a}
uses a MAXSAT solver to compute the maximal set of statements
that may cause the failure given a failed error trace of a
\var{C} program.
\Tool is similar in that we phrase localization as an optimization
problem, allowing the use of ILP to locate the possible cause of a
non-constant-time variable.
However, \Tool focuses on constant-time, which is a relational property, and
hardware which has a substantially different execution model.

\mypara{Synthesizing Assumptions}
%
%
Our approach to synthesizing secrecy assumptions is related to work on
precondition synthesis for memory safety.
Data-driven precondition inference techniques such as~\cite{siLearningLoopInvs,
gargLearningInvariantsUsing2016, ice, padhiDatadrivenPreconditionInference2016a,
siCode2InvDeepLearning2020}, unlike \Tool, require positive and negative
examples to infer preconditions.
\Tool's synthesis technique is an instance of \emph{abductive} inference, which
has been previously used to triage analysis reports by allowing the user to
interactively determine the preconditions under which a program is safe or
unsafe~\cite{dilligAutomatedErrorDiagnosis2012} or to identify the most general
assumptions or context under which a given module can be verified safe
\cite{dilligInductiveInvGen, dilligSynthesisCircularCompositional2017,
calcagnoCompositionalShapeAnalysis2011a,
giacobazziAbductiveAnalysisModular1994}.
Unlike the above, our abduction strategy is tailored
to the relational constant-time property. Furthermore, \Tool
uses information from the verifier to ensure that the
user interaction loop only invokes the ILP solver
(not the slower Horn-clause verifier),
yielding a rapid cycle that pinpoints
the assumptions under which a circuit is constant-time.
In future work, we would like to see, if ideas introduced
in \Tool can be applied to localization, explanation and verification
of other classes of correctness or security properties.

\mypara{Modular Verification of Software and Hardware}
\Tool exploits modularity to verify large circuits by composing
\emph{summaries} of the behaviors of smaller sub-components of
those circuits.
This is a well-know idea in verification; for example, \cite{RHS95} shows how to
perform dataflow analysis of large programs by computing procedure summaries,
and Houdini~\cite{flanaganHoudiniAnnotationAssistant2001} shows how to verify
programs by automatically synthesizing pre- and post- conditions summarizing the
behaviors of individual procedures.
On the hardware side, model checkers like Mocha~\cite{AlurH99b}
and SMV~\cite{McMillan97} use rely-guarantee reasoning to perform
modular verification. Kami~\cite{choiKamiPlatformHighLevel2017}
and \cite{vijayaraghavanModularDeductiveVerification2015}
develops a compositional hardware verification methodology using
the Coq proof assistant. However, the above require the user
to provide module interface abstractions.
There are some approaches that synthesize such abstractions
in an counterexample guided fashion \cite{GuptaMF08, zhangSynthesizingEnvironmentInvariants2020}.
All focus on functional verification of properties of a
\emph{single} run, and do not support abstractions needed to reason about
timing-channels which require relational hyperproperties \cite{hyperprop}.

\bibliographystyle{plain}
\bibliography{references}

\pagebreak

\appendix
\section{Example: Not all Variables Become Public \label{sec:non-pub}}

\mypara{Example 3}
One might think that \Tool requires all variables occurring in branch conditions
to be annotated as public, however, this is not the case.
\Cref{fig:example3} shows an example of such a program.
Running \Tool produces the dependency graph shown in \Cref{fig:ex-3-var-dep}.
\Tool computes root-cause candidates by eliminating constant-time nodes and
edges violating the precedence order. The result is shown in
\Cref{fig:ex-3-root-causes}. Removing all nodes that cannot reach source
\var{out} leaves only nodes \textcolor{BlueViolet}{\var{r3}} and \var{out}, and
since \textcolor{BlueViolet}{\var{r3}} has no predecessors, we identify it as
the earliest node that became non-constant time, and therefore the root cause of
the problem. Solving the ILP constraints yields \textcolor{Emerald}{\var{stall}}
as candidate assumption, and marking \textcolor{Emerald}{\var{stall}} as public and
restarting \Tool verifies constant time execution without the need to mark
\var{cond} as public. This is possible because \Tool is able to prove that
\var{tmp1} and \var{tmp2} have the same colors, irrespective of the value of
\var{cond}, \ie that \colorOf{\var{tmp1}}=\colorOf{\var{tmp2}} holds
irrespective of \var{cond}.   
\begin{lstlisting}[style={verilog-style},label={fig:example3}]
module test(clk, in, cond, bubble, out);
  input wire clk, in, cond, bubble;
  output reg out;
  reg tmp1, tmp2, r2, r3;

  always @(posedge clk) begin
    tmp1 <= in | r3;
    tmp2 <= in & r3;

    if (cond)
      r2 <= tmp1;
    else
      r2 <= tmp2;

    if (stall)
      r3 <= r3;
    else
      r3 <= r2;

    out <= r3;
  end
endmodule
\end{lstlisting}
\begin{figure}[h!]
  \centering
  \resizebox{.5\columnwidth}{!}{
    \begin{tikzpicture}[>=latex,line join=bevel,node distance=0.4cm and 0.6cm]
      \node (n1) [thick,draw,ellipse] {r2 (\redcross, $13$)};
      \node (n3) [thick,draw,ellipse, above right= of n1] {tmp1 (\redcross, $10$)};
      \node (n4) [thick,draw,ellipse, below right= of n1] {tmp2 (\redcross, $10$)};
      \node (n5) [color=BlueViolet,thick,draw,ellipse,  right= of n3] {r3 (\redcross, $7$)};
      \node (n6) [thick,draw,ellipse,  right= of n4] {in (\greencheck, $\bot$)};
      \node (n7) [thick,draw,ellipse,  below right= of n5] {out (\redcross, $10$)};
      \node (n8) [color=Emerald,thick,draw,ellipse, above = of n5] {stall (\greencheck, $\bot$)};
      \node (n9) [thick,draw,ellipse, left = 2.65cm of n8] {cond (\greencheck, $\bot$)};
      
       \draw [->,dashed] (n9) edge  (n1);
       \draw [->,solid] (n3) edge  (n1);
       \draw [->,solid] (n4) edge  (n1);
       \draw [->,solid] (n1) edge [bend left=50] (n5);
       \draw [->,solid] (n5) edge  (n3);
       \draw [->,solid] (n5) edge  (n4);
       \draw [->,solid] (n6) edge  (n3);
       \draw [->,solid] (n6) edge  (n4);
       \draw [->,dashed] (n8) edge  (n5);
       \draw [->,solid] (n5) edge  (n7);

    ;
      \end{tikzpicture}
    }
    \caption{Example 3: Variable dependency graph. 
      }
    \label{fig:ex-3-var-dep}
\end{figure}
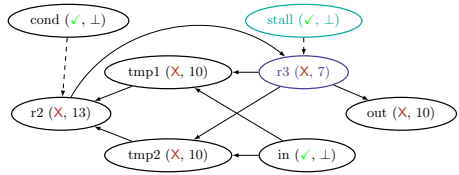
\begin{figure}[h!]
  \centering
  \resizebox{.5\columnwidth}{!}{
    \begin{tikzpicture}[>=latex,line join=bevel,node distance=0.4cm and 0.6cm]
      \node (n1) [thick,draw,ellipse] {r2 (\redcross, $13$)};
      \node (n3) [thick,draw,ellipse, above right= of n1] {tmp1 (\redcross, $10$)};
      \node (n4) [thick,draw,ellipse, below right= of n1] {tmp2 (\redcross, $10$)};
      \node (n5) [color=BlueViolet,thick,draw,ellipse,  right= of n3] {r3 (\redcross, $7$)};
      \node (n7) [thick,draw,ellipse,  below right= of n5] {out (\redcross, $10$)};
      
       \draw [->,solid] (n3) edge  (n1);
       \draw [->,solid] (n4) edge  (n1);
       \draw [->,solid] (n5) edge  (n3);
       \draw [->,solid] (n5) edge  (n4);
       \draw [->,solid] (n5) edge  (n7);

    ;
      \end{tikzpicture}
    }
    \caption{Example 3: Variable dependency graph after eliminating non-ct
      nodes and edges that violate the precedence relation.
      }
    \label{fig:ex-3-root-causes}
\end{figure}
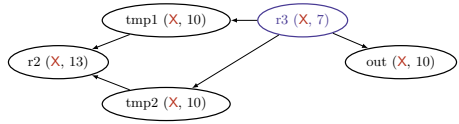

\end{document}